\documentclass[10pt, conference, letterpaper]{IEEEtran}

\usepackage[pdftex]{graphicx}
\usepackage{color}
\usepackage{subfigure}
\usepackage{algorithm}
\usepackage{algorithmic}
\usepackage{listings}
\usepackage{amsmath}
\usepackage{multirow}
\usepackage{url}
\usepackage{tabularx}
\usepackage{booktabs}
\usepackage{paralist} 
\usepackage{balance}

\begin{document}

\title{Cardea: Context--Aware Visual Privacy Protection from Pervasive Cameras}

\author{\IEEEauthorblockN{Jiayu Shu\IEEEauthorrefmark{1},
Rui Zheng\IEEEauthorrefmark{2}, and Pan Hui\IEEEauthorrefmark{3}}
\IEEEauthorblockA{HKUST-DT System and Media Laboratory\\
Hong Kong University of Science and Technology, Hong Kong\\
Email: \IEEEauthorrefmark{1}jshuaa@ust.hk,
\IEEEauthorrefmark{2}rzhengac@ust.hk,
\IEEEauthorrefmark{3}panhui@ust.hk}}

\maketitle

\begin{abstract}
The growing popularity of mobile and wearable devices with built--in cameras, the bright prospect of camera related applications such as augmented reality and life--logging system, the increased ease of taking and sharing photos, and advances in computer vision techniques have greatly facilitated people's lives in many aspects, but have also inevitably raised people's concerns about visual privacy at the same time.
Motivated by recent user studies that people's privacy concerns are dependent on the context, in this paper, we propose Cardea, a context--aware and interactive visual privacy protection framework that enforces privacy protection according to people's privacy preferences.
The framework provides people with fine--grained visual privacy protection using:
\begin{inparaenum}[\itshape i\itshape)]
\item personal privacy profiles, with which people can define their context--dependent privacy preferences; and
\item visual indicators: face features, for devices to automatically locate individuals who request privacy protection; and
\item hand gestures, for people to flexibly interact with cameras to temporarily change their privacy preferences.
\end{inparaenum}
We design and implement the framework consisting of the client app on Android devices and the cloud server.
Our evaluation results confirm this framework is practical and effective with $86\%$ overall accuracy, showing promising future for context--aware visual privacy protection from pervasive cameras.

\end{abstract}

\IEEEpeerreviewmaketitle

\section{Introduction}

Nowadays, built--in cameras have been serving as indispensable components of mobile and wearable devices.
Cameras with smaller size and higher resolution support a number of services and applications such as taking photos, mobile augmented reality, and life--logging systems on devices like smartphones, Microsoft HoloLens \cite{microsoft}, Google Glass \cite{google}, and Narrative Clip \cite{narrative}.
The trend of embedding cameras in wearables will keep growing, an example of which is smart contact lens\footnote{\url{http://money.cnn.com/2016/05/12/technology/eyeball-camera-contact-sony/}}.

However, the ubiquitous presence of cameras, the ease of taking photos and recording video, along with ``always on'' and ``non--overt act'' features threaten individuals' rights to have private or anonymous social lives, raising people's concerns of visual privacy.
More specifically, photos and videos captured without getting permissions from bystanders, and then uploaded to social networking sites, can be accessed by everyone online, potentially leading to invasion of privacy.
Malicious applications on the device may also inadvertently leak captured media data\footnote{\url{http://www.infosecurity-magazine.com/news/popular-android-camera-app-leaks/}}.
What makes it worse is that recognition technologies can link images to specific people, places, and things, thus reveal far more information than expected, making searchable what was not previously considered searchable \cite{shaw2006recognition, acquisti2014face}.
All these possible consequences, whether have been realized by people or not, may hinder their acceptance of advanced wearable consumer products.
A representative example is Google Glass, which has been questioned by 
US Congressional Bi-Partisan Privacy Caucus and Data Protection Commissioners around the world
concerning privacy risks to the public \cite{congress, commission}.
They have huge concerns regarding the privacy of non--users/bystanders, and have raised questions of \textit{``How does Google plan to to prevent Google Glass from unintentionally collecting data about non--users without consent?''} and \textit{``Are product lifecycle guidelines and frameworks, such as Privacy by Design, being implemented in connection with its design and commercialization?''}
From these legal concerns, we are confident that in the future wearable devices with cameras are supposed to implement Privacy by Design before being released to the global markets.
Therefore, we base our research on this assumption and aim to develop the technology that can enable such requirement.

In reality, both legal and technical measurements have been proposed to address privacy issues raised by unauthorized or unnoticed visual information collection.
For instance, Google Glass is banned at places such as banks, hospitals, and bars\footnote{\url{https://www.searchenginejournal.com/top-10-places-that-have-banned-google-glass/66585/}}.
However, prohibiting camera usage does not resolve the issue fundamentally, but instead sacrifices people's rights to capture happy moments even if there is no bystander in the background.
As a result, there are growing needs to design technical solutions to protect individuals' visual privacy in a world where cameras are becoming pervasive.
Some recent attempts are using visual markers such as QR code \cite{bo2014privacy, roesner2014world} or colorful hints like hats \cite{schiff2009respectful} for individuals to actively express their unwillingness to be captured.
However, these visual markers suffer from similar limitations.
First, people are less likely to wear a QR code, despite the technical feasibility of these approaches.
Moreover, privacy concerns vary widely among individuals, and people's privacy preferences change from time to time, following patterns which cannot be conveyed by static visual markers.
In fact, what individuals are doing, with whom, and where at, are all factors which determine whether people think their privacy should be protected.
Therefore, we are looking for a natural, flexible, and fine-grained mechanism for people to express, modify, and control their individualized privacy preferences.

In this paper, we propose a visual privacy protection framework for individuals using:
\begin{inparaenum}[\itshape i\itshape)]
\item personalized privacy profiles, that people can define their context--dependent privacy preferences with a set of privacy related factors including location, scene, and other's presence;
\item face features, for devices to locate individuals who request privacy control; and
\item hand gestures, which help people interact with cameras to temporarily change their privacy preferences.
\end{inparaenum}
By using this framework, the device will automatically compute context factors, compare them with people's privacy profiles, and finally enforce privacy protection conforming to people's privacy preferences.

The rest of the paper is organized as follows: in Section~\ref{sec:practical} we introduce motivation and challenges of practical visual privacy protection; in Section~\ref{sec:overview} we provide a high--level overview of our framework; in Section~\ref{sec:implementation} we describe detailed design and implementation of the system; in Section~\ref{sec:evaluation} we present the evaluation results; in Section~\ref{sec:related-work} we discuss the related work; and finally, in Section~\ref{sec:conclusion} we conclude the paper and discuss plans for future work.

\section{Practical Visual Privacy Protection}
\label{sec:practical}

The goal of our work is to propose an in situ privacy protection approach conforming with the principle of Privacy by Design.
Devices with recording capability can protect bystanders' visual privacy automatically and immediately according to their privacy preferences.
Our work is motivated by the findings from recent user studies, and encounters several challenges that help shape up the final design.

\subsection{Context--dependent Personal Privacy Preferences}
Recently, some works try to understand people's attitudes towards visual privacy concerns raised by pervasive cameras and rapid development of wearable technologies.
Hoyle et al. \cite{hoyle2014privacy, hoyle2015sensitive} find that life--loggers are concerned about the privacy of bystanders, and a combination of factors including time, location, and objects in the photo determines the sensitivity of the photo.
A user study conducted by Denning et al. \cite{denning2014situ} also shows that participants' acceptability of being recorded by Augmented Reality Glasses are dependent on a number of elements, including the place and what they are doing when the recording is taken.
Based on the above findings, we conclude the following points that motivate our work:
\begin{itemize}
\item People's privacy concerns are dependent on \textit{context}. Although \textit{location} is an important factor of privacy concerns,
what individuals are doing and with whom are more essential and crucial factors that directly relate to privacy.
\item People's privacy preferences vary from each other, thus individuals should be able to \textit{express} their own \textit{personal} privacy preferences.
\item People's privacy preferences may change from time to time, therefore individuals need a way to \textit{change} such preferences easily.
\item Generally the public hold positive attitudes towards enforcing privacy protection on images to \textit{respect} others' privacy preferences.
\end{itemize}

\subsection{Challenges, Principles, and Limitations}
According to people's context--dependent privacy concerns and conclusions, a practical visual privacy protection framework is faced with the following challenges:

The first challenge is what elements should be taken into consideration regarding context.
A practical protection approach should seek a balance between granularity and representativity of context, at the same time taking computational complexity into consideration.
In our current design, we choose location, scene, and presence of other people as elements to define context.
Location gives an approximate area scope that requests privacy protection.
Scene describes places, and usually indicates what individuals are doing.
Presence of other people can be a concern when individuals want to keep their social relationships private.

The second challenge is how to inform cameras of bystanders' privacy preferences.
Compared with external marker, individuals' faces are more natural and effective visual cues.
To this end, people need to provide their face features for individual recognition, and then set their personal privacy profiles by selecting context elements they are concerned with.

The third challenge is how people can easily modify their privacy preferences.
Moreover, bystanders should be able to react to cameras immediately when they find being captured without efforts on updating profiles.
Therefore, we offer hand gestures for individuals to ``speak out'' their privacy prefences in the capturing moment.
Once a certain gesture is detected, the image will be processed to retain or remove individual's identifiable information accordingly.

The technical focus of our approach is on proving the feasibility of a context--aware and interactive visual privacy protection framework through system design, implementation, and evaluation.
However, like other technical solutions \cite{halderman2004privacy, schiff2009respectful, jung2014courteous, roesner2014world, aditya2016pic}, our approach only works with compliant devices and users.
Any non--compliant device can still capture images without respecting people's privacy preferences.
Also, we assume the cloud server that stores user profiles is trusted.
Though there are limitations of our approach, we hope our work can motivate more complete visual privacy protection approaches, and finally be integrated 
into default camera subsystems.

\section{Cardea Overview}
\label{sec:overview}

In this section, we first introduce some key concepts.
Then we discuss major components of Cardea.

\subsection{Key Concepts}

\subsubsection*{\textbf{Bystander, Recorder, User, and Cloud}}
A bystander is the person who may be captured by pervasive cameras.
A recorder is the person who holds a device with a built--in camera.
A bystander who worries about his visual privacy can use Cardea to express his privacy preferences.
A recorder who respects bystanders' privacy preferences uses Cardea framework to capture images.
Both the bystander and recorder become users of the privacy protection framework.
The cloud listens to recorder's requests and makes sure captured images are compliant with registered bystander's privacy preferences.


\subsubsection*{\textbf{Privacy Profile}}
Privacy profiles contain context elements, whether enabling hand gestures or not, and the protection action.

\paragraph*{\textbf{a) Context Elements}}
Context elements are factors that reflect people's visual privacy concerns.
Currently, we have considered location, scene, and people in the image.
More elements can be included in the profile to describe people's context--dependent privacy preferences.

\paragraph*{\textbf{b) Gesture Interaction}}
We define ``Yes'' and ``No'' gestures.
A ``Yes'' gesture is a victory sign, which means \textit{``I would like to appear in the photo.''}
A ``No'' gesture is a palm, which represents \textit{``I do not want to be photographed.''}
They are consistent with people's usual expectations, and is commonly used in daily lives to express willingness or unwillingness to be captured by others.

\paragraph*{\textbf{c) Privacy Protection Action}}
This action refers to removal of identifiable information.
In our implementation, we blur user's face as an example of protection action.
Other methods such as replacing the face with an average face, blurring the whole body, or blending the body region into the background can be integrated into our framework.

\subsection{System Overview}
\begin{figure}[!t]
\centering
\includegraphics[width=0.9\columnwidth]{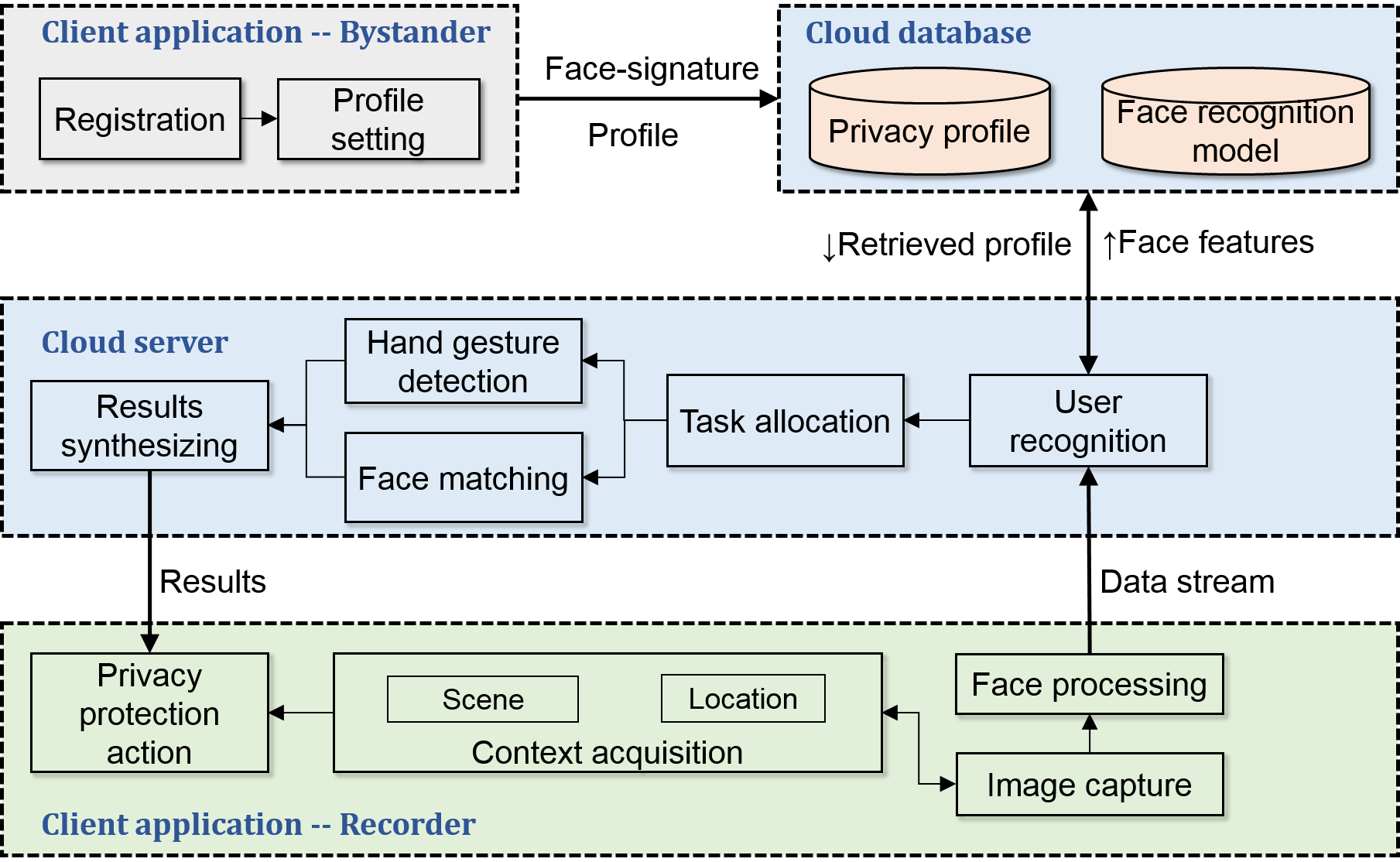}
\caption{Cardea overview.
The client application allows bystanders to register and set privacy profile.
Users can use the platform to take pictures while respecting others privacy preferences.
The cloud server recognizes users, retrieves privacy profiles, and performs necessary computations.}
\label{fig:architecture}
\end{figure}

Cardea is composed of the client app and cloud server.
It works based on data exchange and collaborative computing of both the client and cloud sides.
The major components and interactions are shown in Figure~\ref{fig:architecture}.

\textbf{\textit{Cardea Bystander application:}}
Bystanders use Cardea application to register as users and define their privacy profiles.

A bystander is presented with the user interface of Cardea client application for registration.
The application will extract face features from the bystander automatically, and then upload to the cloud to train the face recognition model.
After registration, a user can define his context--dependent privacy profile.
The profile will also be sent to the cloud and stored in the cloud database for future retrieval.
Both features and profiles can be updated.

\textbf{\textit{Cardea Recorder application:}}
Recorders use Cardea application to take images.
After capturing an image, context elements will be computed locally on the device.
Meanwhile, the application detects faces and extracts face features.
As hand gesture recognition is extremely computationally--intensive, the task will be offloaded to the cloud.
To lower the risk of privacy leakage during transmission, detected faces will be removed from the image before sending the compressed image data and face features to the cloud.

Upon receiving synthesized results from the cloud, the final privacy protection action will be performed on the image according to the computing results from both the local device and remote cloud.
Details of how to make privacy protection decisions are discussed in Section~\ref{sec:implementation}.

\textbf{\textit{Cardea Cloud Server:}}
The function of the cloud server is twofold:
\begin{inparaenum}[\itshape a\itshape)]
\item storing users' profiles and training the user recognition model; and
\item responding to clients' requests by recognizing users in the image and perform corresponding tasks.
\end{inparaenum}

When the cloud server receives requests from the client application, it first recognizes users in the image using the face recognition model.
If there is any recognized user, the corresponding privacy profile will be retrieved to trigger related tasks.
For example, if ``Yes'' or ``No" gesture is enabled, gesture recognition task will start.
Finally, computing results on the cloud will be synthesized and sent back to the client.

\section{Design and Implementation}
\label{sec:implementation}

\begin{figure}[!t]
\centering
\includegraphics[width=0.95\columnwidth]{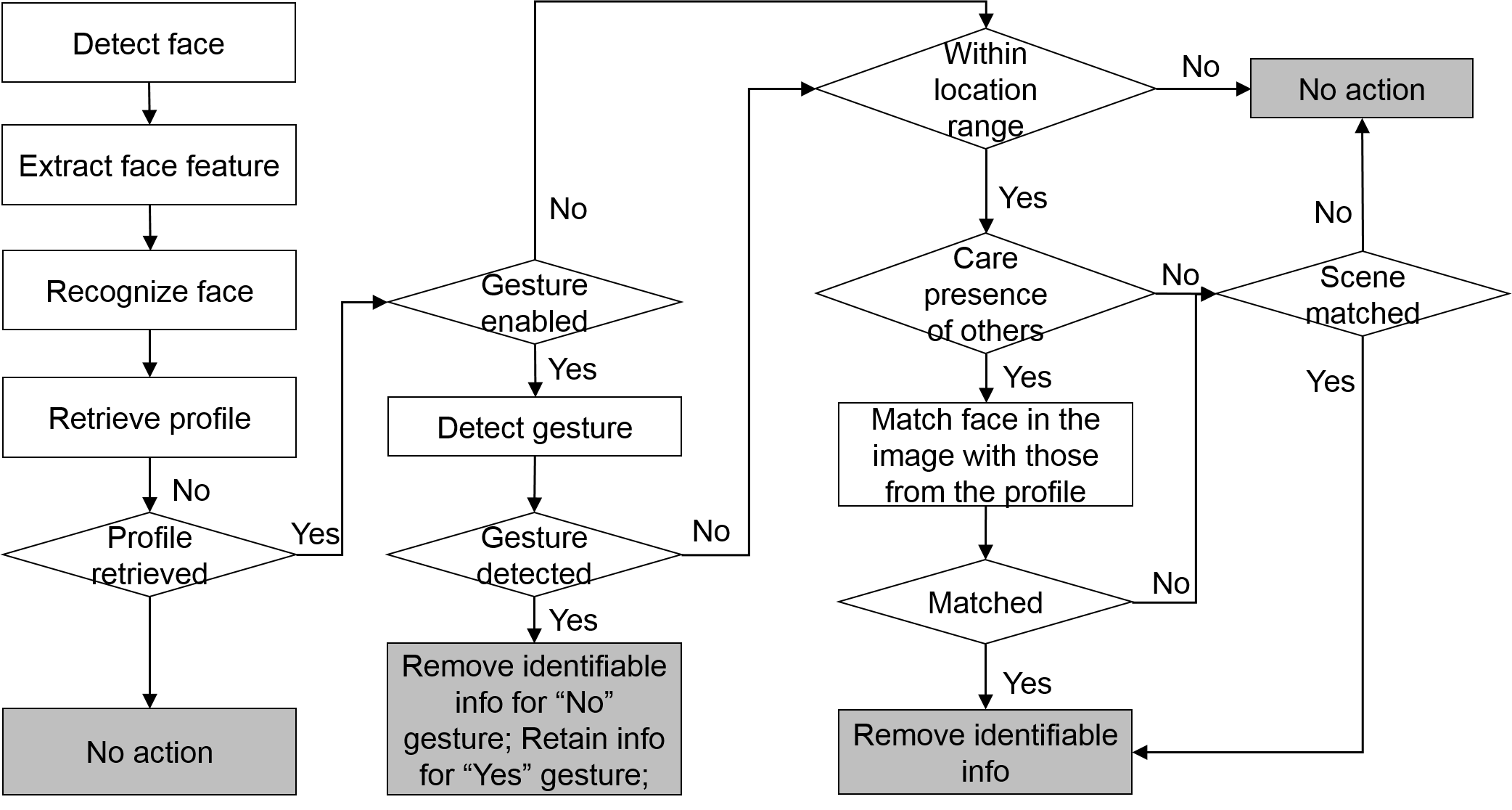}
\caption{Workflow of Cardea.
Privacy protection decision is made according to computing results from local recording device and remote cloud.
Final privacy protection action is performed on the original image.}
\label{fig:workflow}
\end{figure}

The client application and cloud server are implemented on Android smartphones and desktop computer respectively.
Next, we discuss the whole decision making procedures as shown in Figure~\ref{fig:workflow}, which involves both the client application and cloud server.

After an image is captured using Cardea, we first detect faces and extract face features.
If there is any face in the image, we then input the extracted face features into the pre--trained face recognition model to recognize users.
If any Cardea user is recognized, his privacy profile will be retrieved.
Otherwise, we do nothing to the original image.

The profile describes whether the user enables hand gestures, and the location where he is concerned about his visual privacy.
The profile may also contain information about selected scenes and other people that the user especially cares about.
Among these factors, ``Yes'' and ``No'' gestures have the highest priority, that is, if any gesture is detected and enabled in user's profile, privacy protection action will be determined instantly.
In this way, 
users can temporarily modify their privacy preferences for cases in which that they are aware of being photographed, without updating their privacy profiles.

If gestures are not enabled or not detected for a user, and the image is captured within the area/location user specifies, we will start tasks of recognizing scene and finding if there is anyone in the image that the user cares about according to his privacy profile.
When all the tasks are finished, the final privacy protection decision will be made and performed on the original image automatically to protect user's visual privacy.

\subsection{User Recognition}
User recognition is to identify users that request visual privacy protection in the image.

\subsubsection*{\textbf{Face Detection}}
Face detection is to detect face regions in the whole image.
It is the first step for all face image processing.
We use Adaboost cascade classifier \cite{viola2004robust} implemented in OpenCV \cite{opencv} library to detect face regions, which has real--time performance.
We then filter detected face regions using Dlib \cite{dlib} to reduce false positives.

\subsubsection*{\textbf{Face Feature Extraction}}
Face feature extraction is to find a compact yet discriminative representation that can best describe a face instead of using raw pixels.
Convolutional Neural Networks (CNNs) have achieved state-of-the-art results on many computer vision tasks, including face recognition, which has been studied for years using different image processing techniques \cite{zhao2003face, learned2016labeled}.
Compared with conventional features such as Local Binary Patterns \cite{ahonen2006face}, features extracted from the CNN models yield much better performance.
Therefore, we use 256--dimensional CNN--features extracted from a lightened CNN model \cite{wu2015lightened}, which has a small size, fast speed of feature extraction and low-dimensional representation.
The model is pre-trained with the CASIA-WebFace dataset containing $\sim 0.5M$ face images from 10575 subjects \cite{casia}.
We use the $1KB$ output of the ``\textit{eltwise\_fc1}'' layer from the CNN model as face features.
We have also experimented a deeper CNN model with VGG architecture \cite{parkhi2015deep}, which is widely used in CNN models.
However, the computational burden 
using VGG model is too heavy compared with the lighted CNN model, with no obvious performance improvement.
We use the open source deep learning framework Caffe \cite{jia2014caffe} and its Android library \cite{caffeandroid} to extract face features using the lightened CNN model on Android smartphone,

\subsubsection*{\textbf{Face Recognition}}
Face recognition is to identify users in the image.
We train the face recognition model using Support Vector Machine (SVM), which 
can achieve good results without much training data.
The model training is a supervised learning process.
The input training data for the face recognition model are face features from Cardea users.
We train the SVM model using LibSVM \cite{chang2011libsvm} with linear kernel, as the number of training samples from each user is smaller than the dimensions of each input data.

With the face recognition model, we can get the prediction result of the input face feature vector.
The output is $\{p_i\}, i \in 1, 2, \ldots, n$, where $p_i$ is the probability of being user $i$.
We set a probability threshold $T_p$, and only when $\max_{i}{p_i} \ge T_p$ does it mean the user is recognized.
This is to avoid the cases when the input face feature from a non--registered bystander is mistakenly recognized as a registered user.
The threshold $T_p$ is chosen based on experiments that will be described in the next section.

\subsection{Context--aware Computing}
Location, scene, and people in the image are three context elements defined in users' privacy profiles.
We acquire these context elements in different ways after the image is captured.
They decide the final privacy protection action to be performed on the raw image.

\subsubsection*{\textbf{Location Filtering}}
Location provides coarse control for individuals.
Users can name a concrete sensitive area in which that they may have privacy worries.
By specifying locations, such as a campus, user's privacy control will work in the area of the campus.
We directly obtain the location from the GPS when the image is taken.

\subsubsection*{\textbf{Scene Classification}}

\begin{table}[tb]
\centering
\caption{Ten general scene groups}
\label{tab:scene}
\begin{tabularx}{0.46\textwidth}{lcl}
\toprule
\textbf{Group name}      & \textbf{Scenes \#}     & \textbf{Examples}   \\ \midrule
Shopping        & $20$            & clothing store, market, supermarket\\ \midrule
Travelling      & $9$             & airport, bus station, subway platform\\ \midrule
Park \& street  & $12$            & downtown, park, street, alley\\ \midrule
Eating \&       & $18$            & bar, bistro, cafeteria, coffee shop,\\
drinking        &                 & fastfood restaurant, food court\\ \midrule
Working \&      & $9$             & classroom, conference center, library,\\
study           &                 & office, reading room\\ \midrule
Scantily clad   & $12$            & beach, swimming pool, water park\\ \midrule
Medical care    & $2$             & hospital room, nursing home\\ \midrule
Religion        & $11$            & cathedral, chapel, church, temple\\ \midrule
Entertainment   & $5$             & amusement park, ballroom, discotheque\\ \midrule
\textit{All}    & $\mathit{98}$   &   \\ \bottomrule
\end{tabularx}
\end{table}

Scene context is a complex concept that not only relates to places, but also gives clues about what people may be doing.
We summarize $9$ general scene groups.
In this way, people can select the general scene group, instead of listing all places they care about.
We fine--tune the pre--trained CNN model to do the scene classification.
The detailed procedures are described below.

\paragraph{Data Preparing and Preprocessing}
The data for training scene classification is from Places2 dataset \cite{places2dataset}.
At the time when Cardea is built, Places2 dataset contained $401$ categories with more than $8$ millions training images.
Among $401$ scene categories, we choose $98$ categories, and then group them into $9$ general scene groups as listed in Table~\ref{tab:scene} based on their contextual similarity.
The grouped scenes are either pervasive or common scenes in daily life that may have a number of bystanders, or places where people are more likely to have privacy concerns.
This scene subset is composed of $1.9$ million training images and $4900$ validation images, $50$ validation images for each category.

\paragraph{Training}
The training procedure is a standard fine--tuning process.
We first extract features of all training images from $98$ categories.
The features are the output of ``\textit{fc7}'' layer using the pre--trained AlexNet model provided by Places2 project as a feature extractor \cite{places2dataset}.
With all features, we then train a Softmax classifier for $98$ scene categories.

The category classifier achieves $55\%$ validation accuracy on $98$ categories.
There is no other benchmark result on the subset we choose, but recent benchmarks give $53\% \sim 56\%$ validation accuracy on the new Places2 dataset with 365 categories \cite{places2dataset}.
Both feature extraction and classifier training are implemented using Caffe library \cite{jia2014caffe, caffe}.


\paragraph{Prediction}
To get the probability of each scene group, we simply sum up the probabilities of categories in the same group.
The final prediction result is the group that has the highest probability among $9$ groups, which can be seen as a hard--coded clustering process.


The prediction accuracy of our scene group model on the validation set achieves $83\%$.
As category prediction probabilities are usually distributed among similar categories that belong to the same group, the group prediction results are superior to category prediction results.
It is what we desire for the purpose of privacy protection based on more general scene context description.

\subsubsection*{\textbf{Presence of Other People}}
The third context element we take into account is people in the image.
For example, user \textit{Alice} can upload $10$ face features of \textit{Bob} with whom \textit{Alice} does not want to be photographed.
The task is to determine whether \textit{Bob} also appears in the image when \textit{Alice} is captured in the image.

A simple similarity matching is adopted as the number of potential matches (i.e., people in the image except \textit{Alice} herself) to be considered is small.
We apply cosine similarity as distance metric to get the distance value between a pair of feature vectors.
Assume we have $10$ of \textit{Bob's} face features and $1$ face feature extracted from a person \textit{P} in the image.
We compare \textit{P's} feature with each of \textit{Bob's} features.
If the value does not exceed the distance threshold $T_d$, we regard it as a hit.
When the hit ratio of \textit{Bob's} features reaches the ratio threshold $T_r$, we believe \textit{P} is \textit{Bob}, therefore \textit{Alice} should be protected.
Detailed face matching method is described in Algorithm~\ref{alg:face}.

\begin{algorithm}[tbp]
\caption{Face Matching}
\label{alg:face}
\begin{algorithmic}[1] 
\begin{scriptsize}
\STATE initialize \textit{P's} feature $f_0$, \textit{Bob's} feature $f_{i}, i \in 1, \ldots, N$, distance threshold $T_d$, hit ratio threshold $T_r$ 
\STATE $m \gets 0$ $//$ number of hits
\FOR{$i=1$ to $N$}
	\STATE $d_{i} \gets dis(f_{0},f_{i})$ $//$ $dis(x,y)$ returns the distance between $x$ and $y$
	\IF{$d_{i} \leq T_{d}$}
		\STATE $m \gets m + 1$
 	\ENDIF
\ENDFOR
\IF{$m / N \geq T_{r}$}
    \RETURN true $//$ \textit{P} is \textit{Bob}
\ELSE 
    \RETURN false $//$ \textit{P} and \textit{Bob} are two persons
\ENDIF
\end{scriptsize}
\end{algorithmic}
\end{algorithm}

\subsection{Interaction using Hand Gestures}
Hand gestures are natural body languages.
Our goal is to detect and recognize ``Yes'' and ``No'' gestures in the image.
However, hand gesture recognition in images taken by regular cameras with cluttered background is a difficult task.
Conventional skin color--based hand detectors 
will fail dramatically in complex environments.
In our design, we use the state--of--the--art object detection framework Faster Region--based CNN (Faster R-CNN) \cite{ren2015faster} to train a hand gesture detector as described below.

\subsubsection*{\textbf{Gesture Recognition Model Training}}
According to the gesture recognition task, we categorize hand gestures into $3$ classes:
\begin{inparaenum}[\itshape 1\itshape)]
\item ``Yes'' gesture;
\item ``No'' gesture; and
\item ``Natural'' gesture, which is a hand in any other pose.
\end{inparaenum}
The date used to train the gesture recognition model is composed of $13050$ ``Natural'' gesture instances from $5628$ images in VGG hand dataset \cite{vgghanddataset, mittal2011hand}, and $4712$ ``Yes'' gesture instances, $3503$ ``No'' gesture instances from $5900$ images crawled from Google and Flickr.
Each annotation consists of an axis aligned bounding rectangle, along with its gesture class.


With the comprehensive gesture dataset and VGG16 pre--trained model provided by Faster--RCNN library, we fine--tune the ``\textit{conv3\_1}'' and upper layers, together with region proposal layers and Fast--RCNN detection layers \cite{girshick2015fast}.
The detailed training procedures can be found in \cite{fasterrcnnpython}.
After the gesture is detected, we link it to the nearest face, which requires users to show their hands near their faces when using gestures.


\begin{figure*}[tbf]
\centering
\subfigure[Cardea client application user interface]{
    \includegraphics[width=0.48\textwidth]{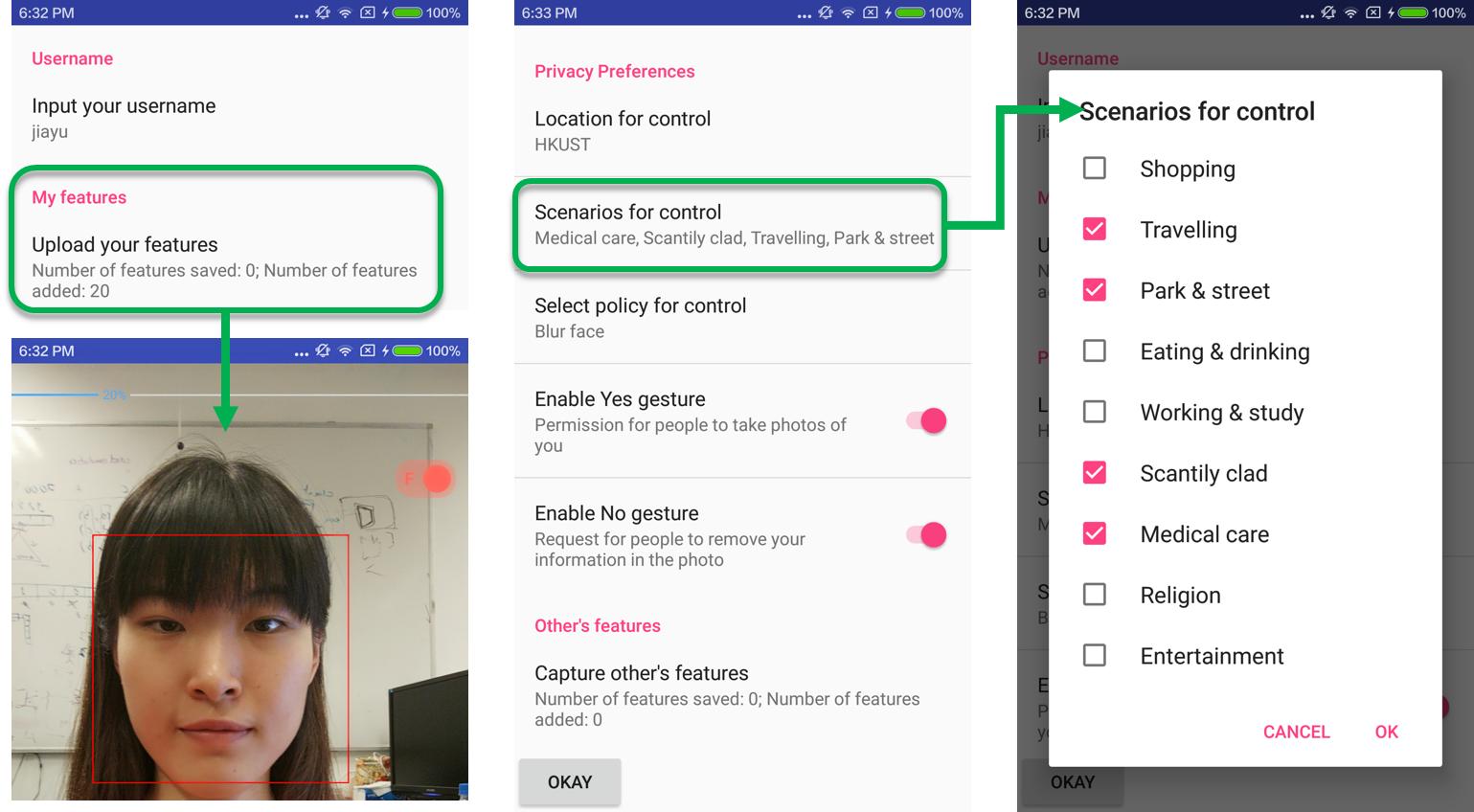}
    \label{fig:ui}
    }
\hspace{-0.1cm}
\subfigure[Privacy protection example]{
    \includegraphics[width=0.48\textwidth]{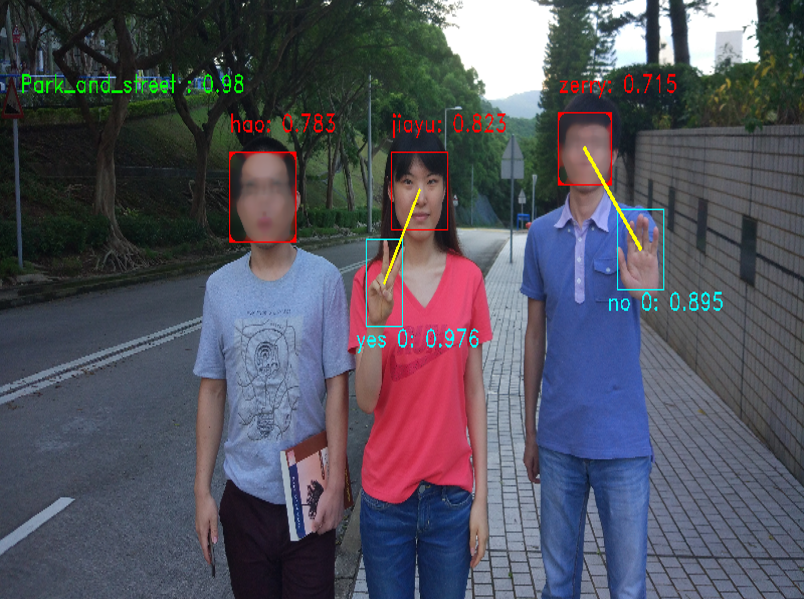}
    \label{fig:result}
    }
\caption{Cardea user interface and privacy protection results.
In (a), \textit{jiayu} registers as a Cardea user by extracting and uploading her $20$ face features.
She specifies \textit{HKUST}, four scene groups for privacy protection.
She also enables ``Yes'' and ``No'' gestures.
In (b), a picture is taken in \textit{HKUST}.
$3$ registered users, one ``Yes'' and one ``No'' gesture are recognized.
The scene is correctly predicted as \textit{Park \& street}.
Therefore, \textit{jiayu's} face is not blurred due to her ``Yes'' gesture.
Prediction probabilities are also shown in (b).}
\label{fig:ui-result}
\end{figure*}

We show the user interface of Cardea client application in Figure~\ref{fig:ui}, together with an example of image processing results in Figure~\ref{fig:result}.

\section{Evaluation}
\label{sec:evaluation}

In this section, we present evaluation results along $3$ axes:
\begin{inparaenum}[\itshape 1\itshape)]
\item \textit{\textbf{vision micro--benchmarks}}, to evaluate the performance of different computer vision tasks, including face recognition and matching, scene classification, and hand gesture recognition.\\
\item \textit{\textbf{system overall performance}}, according to final privacy protection decisions and users' privacy preferences.\\
\item \textit{\textbf{runtime and energy consumption}}, which shows the processing time of each part, and energy consumed on one image.\\
\end{inparaenum}


\subsection{Accuracy of Face Recognition and Matching}
We first select $50$ subjects from LFW dataset \cite{lfw} who has more than $10$ images as registered users.
Note that subjects in the CASIA WebFace Database used to train the lightened CNN model do not overlap with those in LFW \cite{yi2014learning}.
For each subject, we extract at least $100$ face features form Youtube video to simulate the process of user registration.
In total, we get $5042$ feature vectors.
These features are then divided into the training set and validation set.
In addition, we collect user test set and non--user test set to evaluate the face recognition accuracy.
The user test set is composed of $511$ face feature vectors from online images of all $50$ registered users.
The non--user test set consists of $166$ face feature vectors from $100$ subjects in LFW database whose names start with ``\textit{A}'' as non--registered bystanders.

\begin{figure}[tbp]
\centering
\subfigure[Training accuracy]{
  \includegraphics[width=0.48\columnwidth]{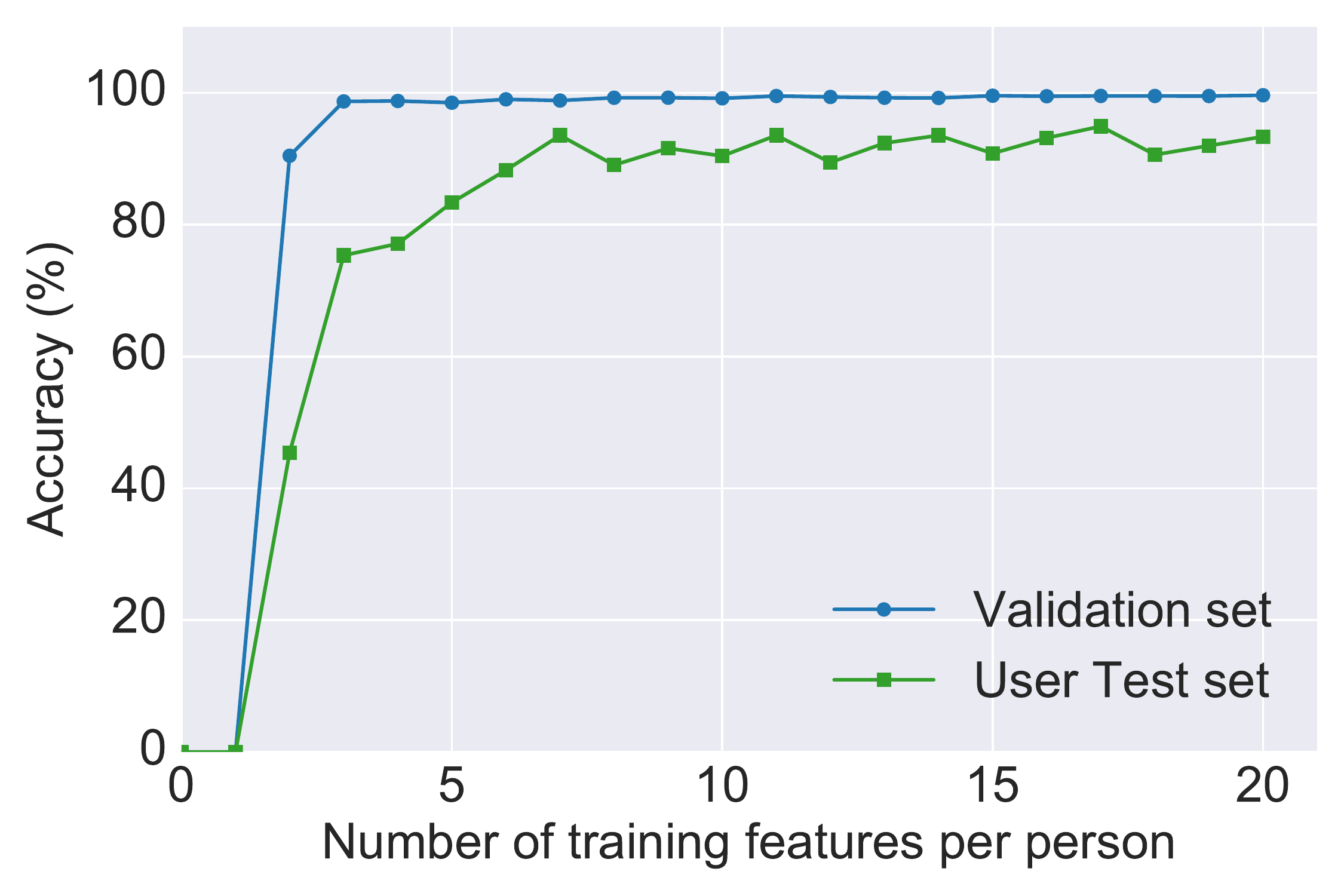}
  \label{fig:accnum}
  }
\hspace{-0.4cm}
\subfigure[Testing accuracy with threshold $T_p$]{
  \includegraphics[width=0.48\columnwidth]{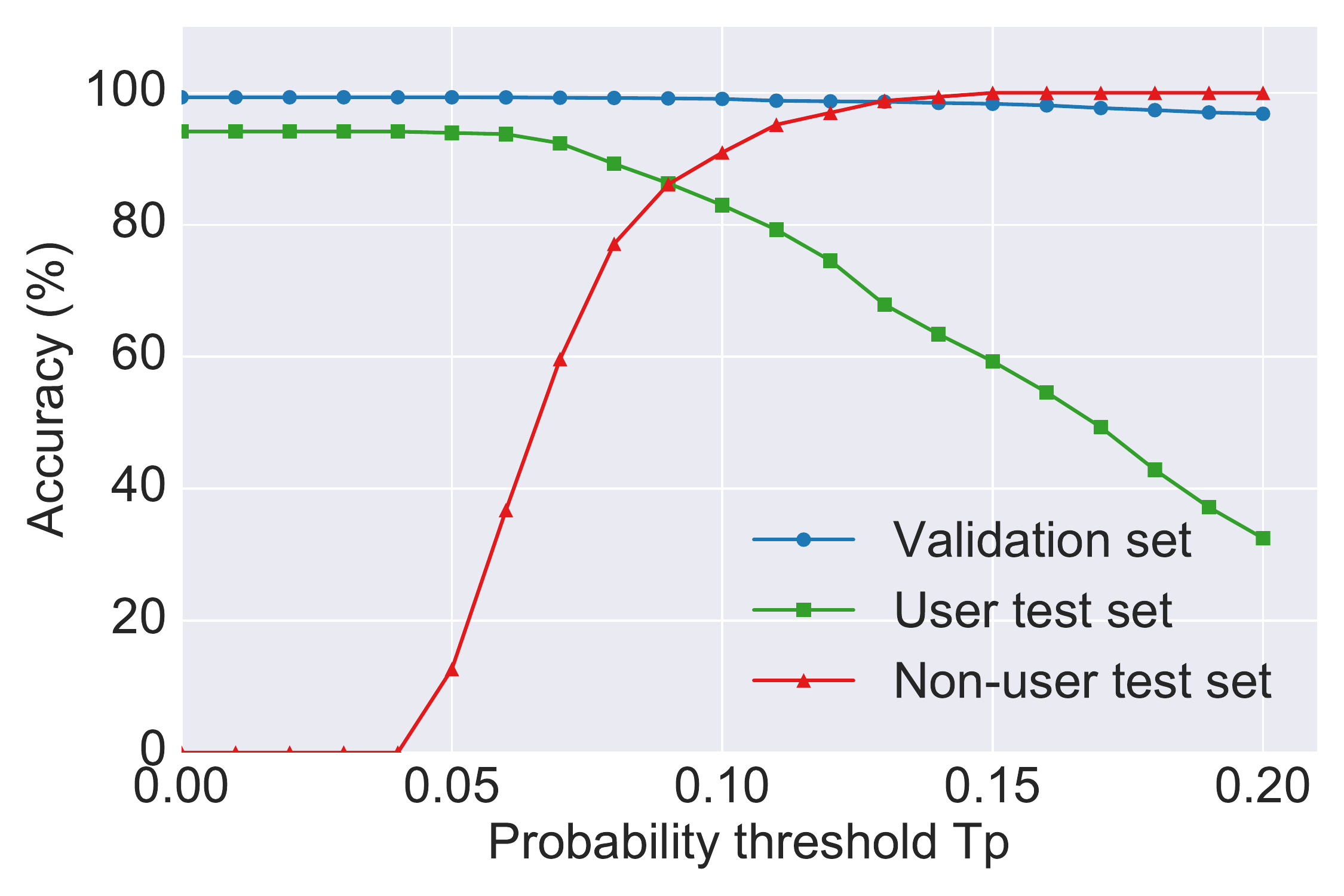}
  \label{fig:accthreshold}
  }
\caption{Face recognition accuracy.}
\label{fig:face-recognition}
\end{figure}

Figure~\ref{fig:accnum} shows the recognition accuracy as to validation set and user test set with different numbers of features per person used for training.
The accuracy refers to the fraction of faces that are correctly recognized.
The results show the model trained with $10 \sim 20$ features per person can achieve near $100\%$ accuracy on the validation set and over $90\%$ accuracy on the user test set.
Little improvement can be achieved with more training data.
Therefore, we train the face recognition model with $20$ features per person.
Figure~\ref{fig:accthreshold} shows the overall accuracy with probability threshold $T_p$.
For non--user test set, the accuracy means the fraction of faces that are not recognized as registered users.
As a result, the accuracy will increase for the non--user test set but decrease for the validation set and user test set when $T_p$ goes up.
To make sure that registered users can be correctly recognized, and non-registered users will not be mistakenly recognized, we choose $T_p$ to be $0.08 \sim 0.09$, which achieves over $80\%$ recognition accuracy for both users and non--users.

\begin{figure}[tbp]
\centering
\subfigure[Cosine distance]{
  \includegraphics[width=0.48\columnwidth]{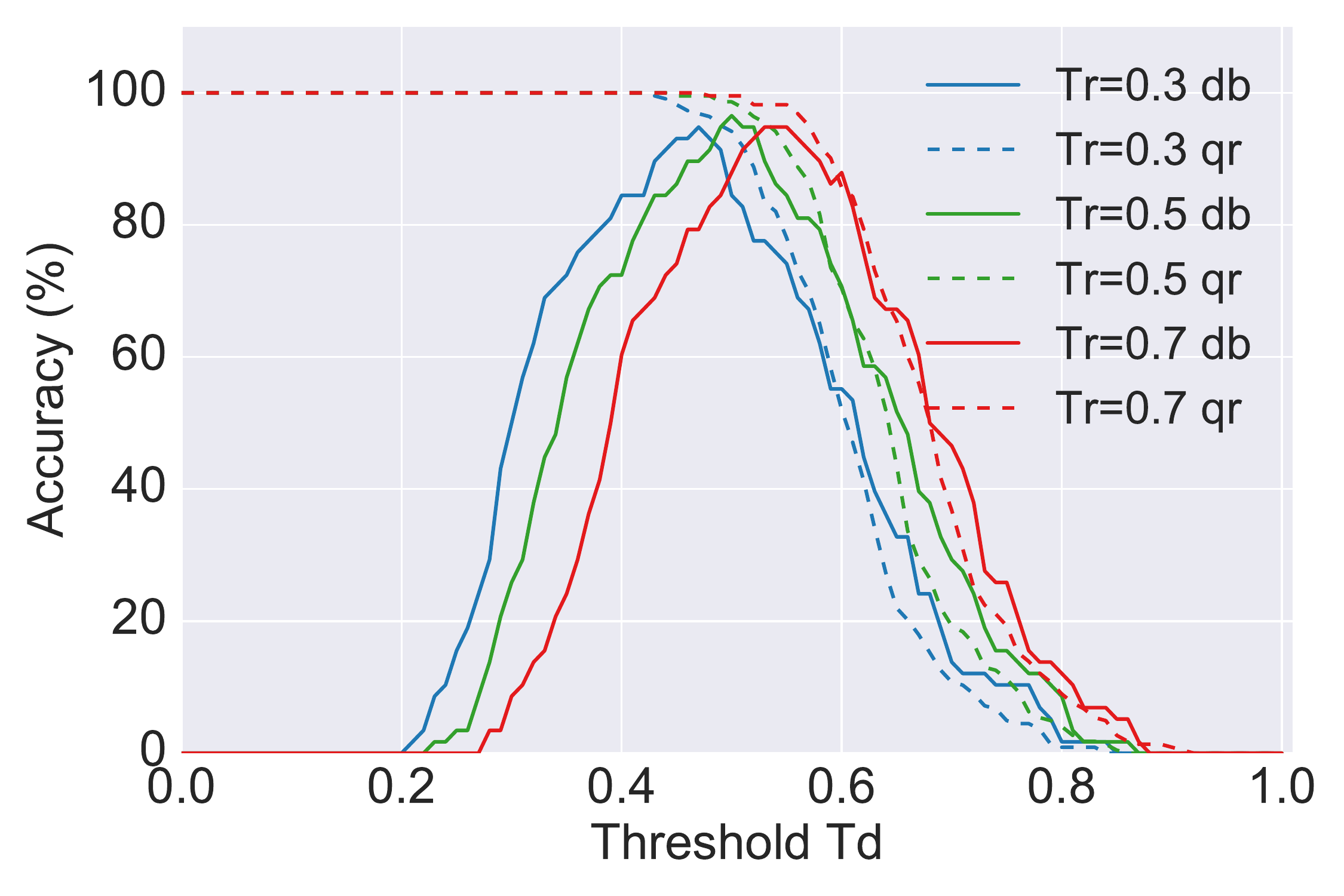}
  \label{fig:acccosine}
  }
\hspace{-0.4cm}
\subfigure[Euclidean distance]{
  \includegraphics[width=0.48\columnwidth]{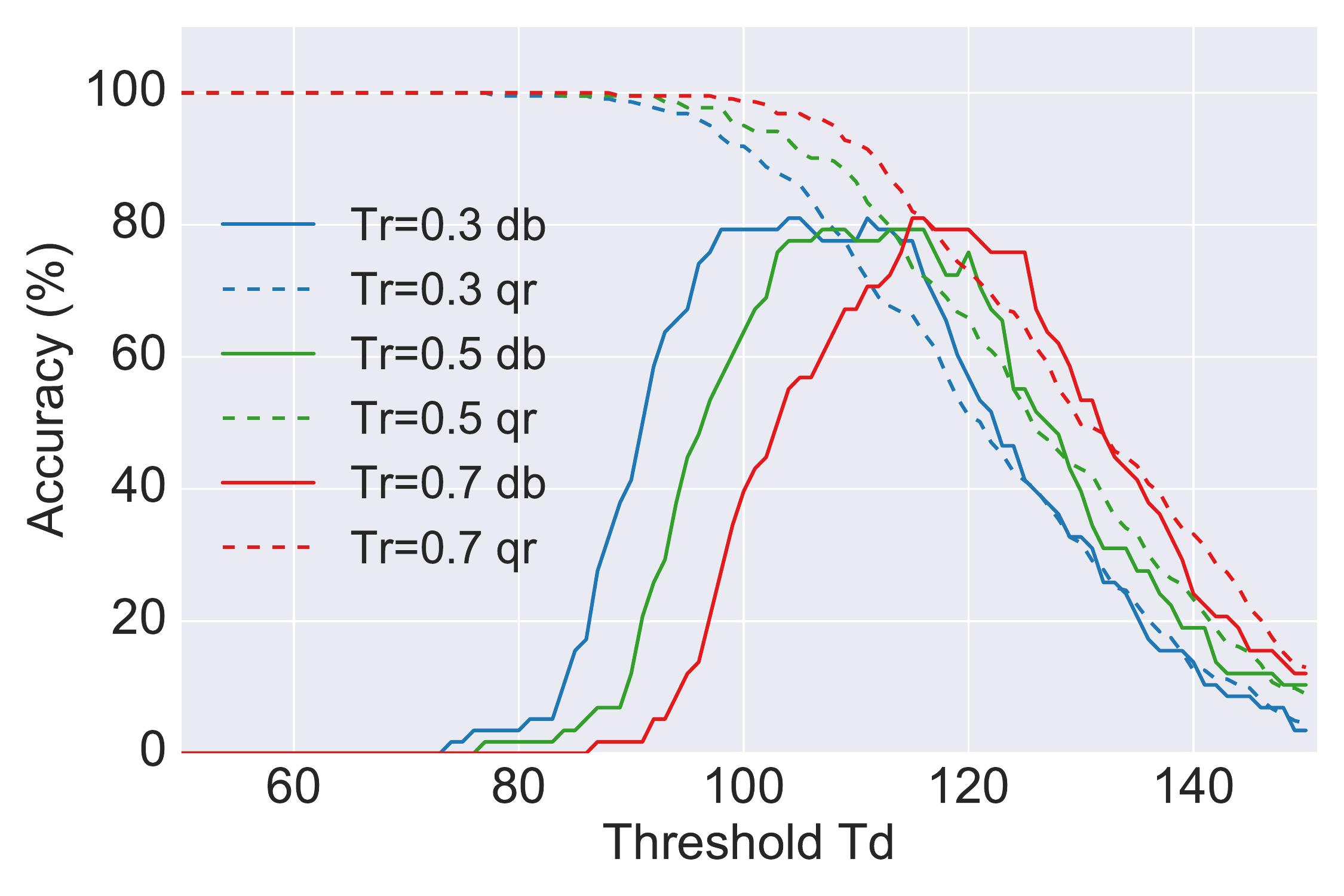}
  \label{fig:accl2}
  }
\caption{Face matching accuracy with different distance threshold $T_d$ and ratio threshold $T_r$.}
\label{fig:face-matching}
\end{figure}

To evaluate face matching performance, we still use the user test set.
Subjects who have more than $20$ features are regarded as the database group, the rest are the query group.
Similar to face recognition accuracy, we break face matching accuracy into two parts: for persons belonging to the database group, we need to correctly match features only to the correct person; for persons not belonging to the database group, we should not match features to anyone.
Figure~\ref{fig:face-matching} shows the matching accuracy with Cosine distance and Euclidean distance respectively.
The results show that Cosine distance can achieve better performance with near $100\%$ accuracy for both situations in which people belong to the database or query group.
The preferable parameters would be distance threshold $T_d \approx 0.5$, and ratio threshold $T_r \approx 0.5$.

Overall, the face recognition and matching methods we employ with appropriate thresholds $T_p, T_d,$ and $T_r$ can effectively and efficiently recognize users and match faces in the images. More importantly, only a small number of features are needed for training the face recognition model and for face matching algorithm.

\subsection{Performance of Scene Classification}

\begin{figure}[tbp]
\centering
\subfigure[Recall]{
  \includegraphics[width=0.8\columnwidth]{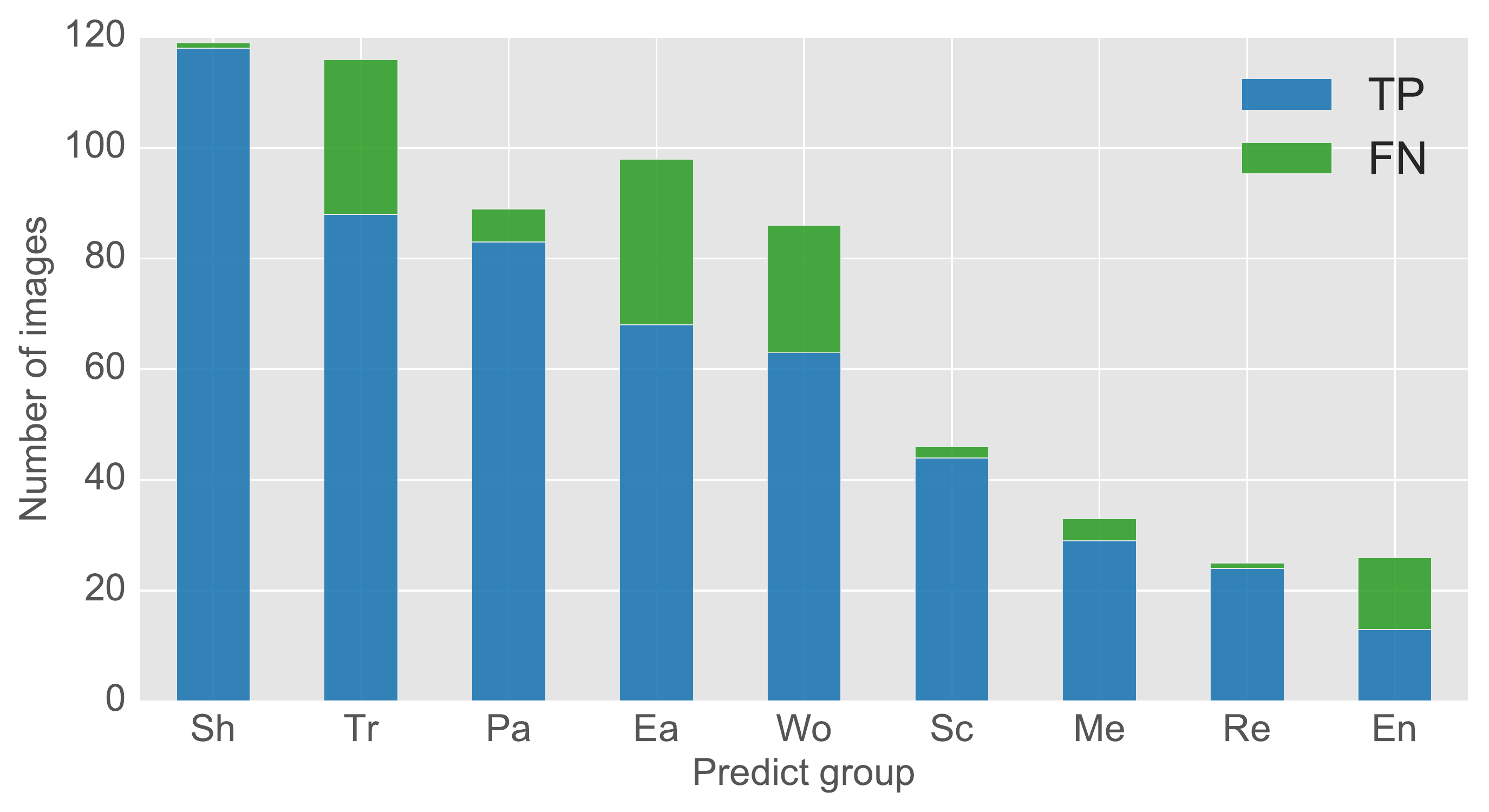}
  \label{fig:scenerecall}
  }
\subfigure[Confusion matrix]{
  \includegraphics[width=0.8\columnwidth]{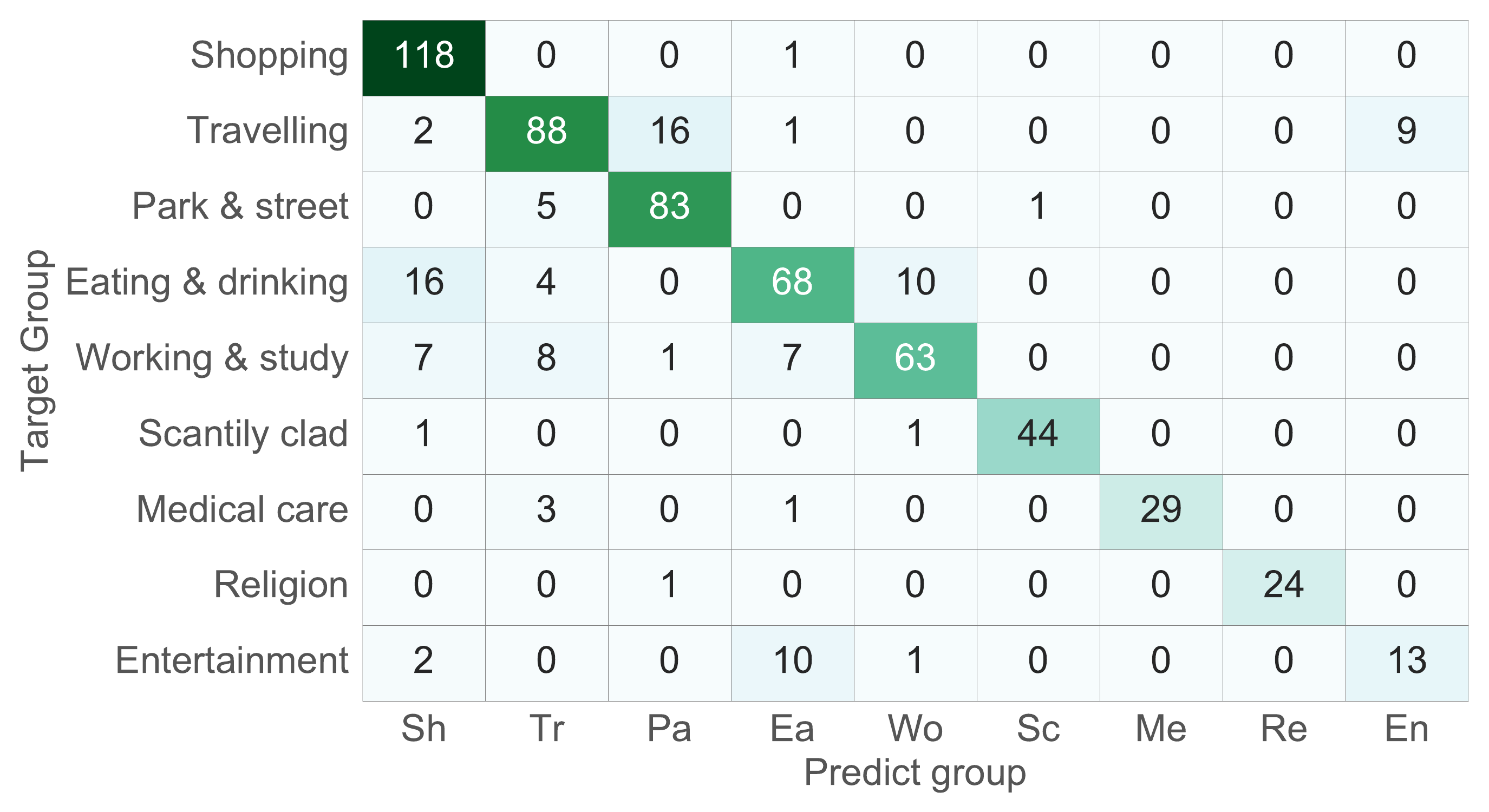}
  \label{fig:sceneconfusion}
  }
\caption{Scene classification results.}
\label{fig:scene-classification}
\end{figure}

We recruited $8$ volunteers and asked them to take pictures ``in the wild'' belonging to $9$ general scene groups.
After manually annotating these pictures, we had $759$ images in total, with $638$ images in $9$ scene groups we are interested in.
We then predict scene group for each image using the scene classification model we trained.

The number of images and classification result of each group are shown in Figure~\ref{fig:scenerecall}.
The true positives (TP) refers to images that are correctly classified, and false negatives (FN) are those classified as other scene groups.
Overall, we can achieve {$0.83$} recall (TP/(TP+FN)), and $5$ scene groups are with recall more than $0.90$.
It is worth mention that the recall of scene groups \textit{Scantily clad (Sc)}, \textit{Medical care (Me)}, and \textit{Religion (Re)} exceed $0.95$, which provides strong support to protect users' privacy in sensitive scenes.

We also give the detailed classification confusion matrix in Figure~\ref{fig:sceneconfusion}.
It shows that most FN of \textit{Eating \& drinking (Ea)} are classified as \textit{Shopping (Sh)} or \textit{Working \& study (Wo)}, and most FN of \textit{Wo} are classified as \textit{Ea} or \textit{Sh}.
The reason is that boundaries between \textit{Sh}, \textit{Ea}, or \textit{Wo} are not clear.
For example, shopping malls have food courts, or people study in coffee shop.
The same reason accounts for the confusion between \textit{Park \& street (Pa)} and \textit{Travelling (Tr)}.
Moreover, for scene categories such as pub and bar, people may group them into \textit{Ea} or \textit{En}.
Therefore, a safe way is to select more scene groups, for instance, both \textit{Ea} and \textit{En} when you go to a pub at night.

In general, the evaluation results from images captured ``in the wild'' demonstrate that most of scenes can be correctly classified, the performance is especially satisfactory for those sensitive scenes.


\subsection{Performance of Gesture Recognition}

\begin{figure*}[tbp]
\centering
\subfigure[Recall for different scenes]{
  \includegraphics[width=0.31\textwidth]{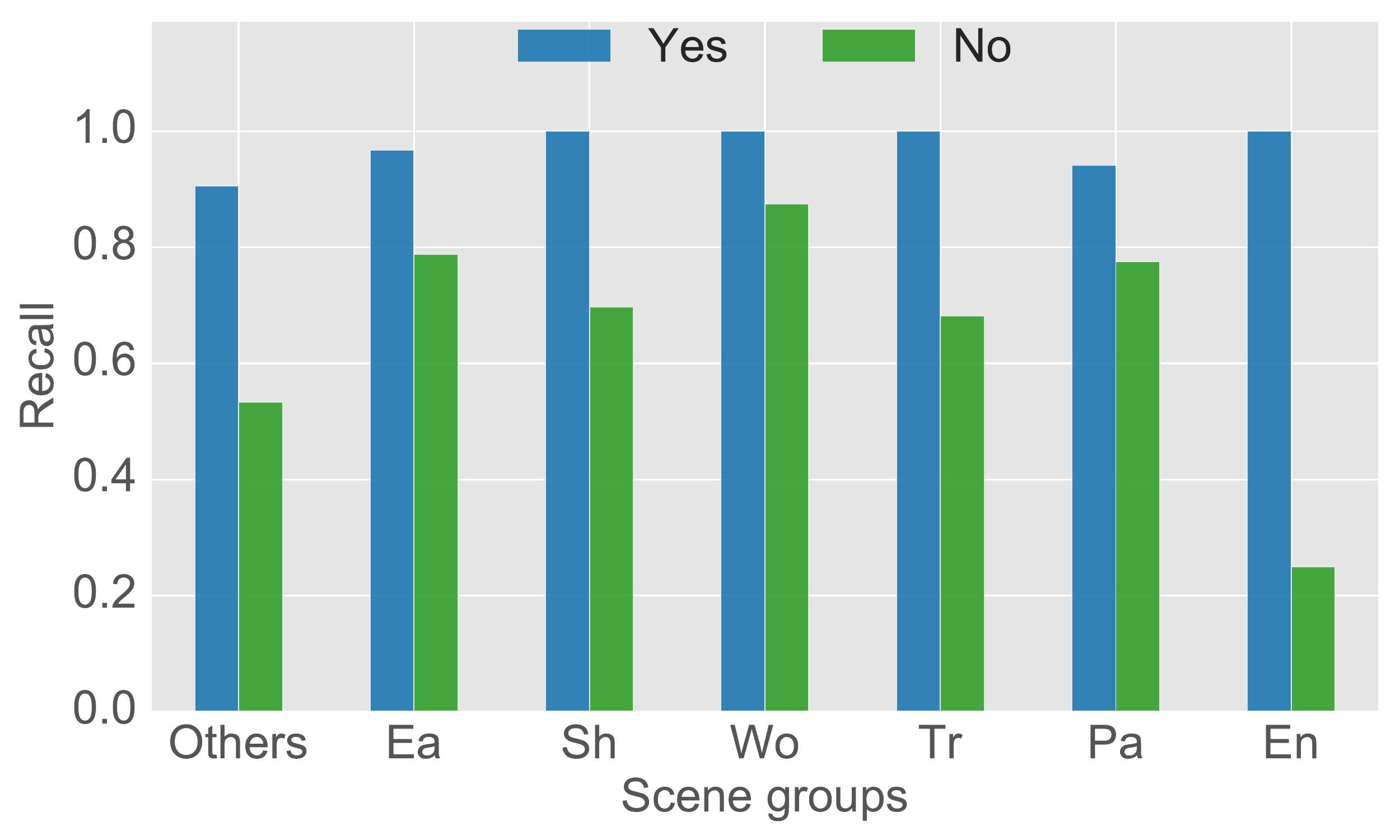}
  \label{fig:handcall}
  }
\subfigure[precision for different scenes]{
  \includegraphics[width=0.31\textwidth]{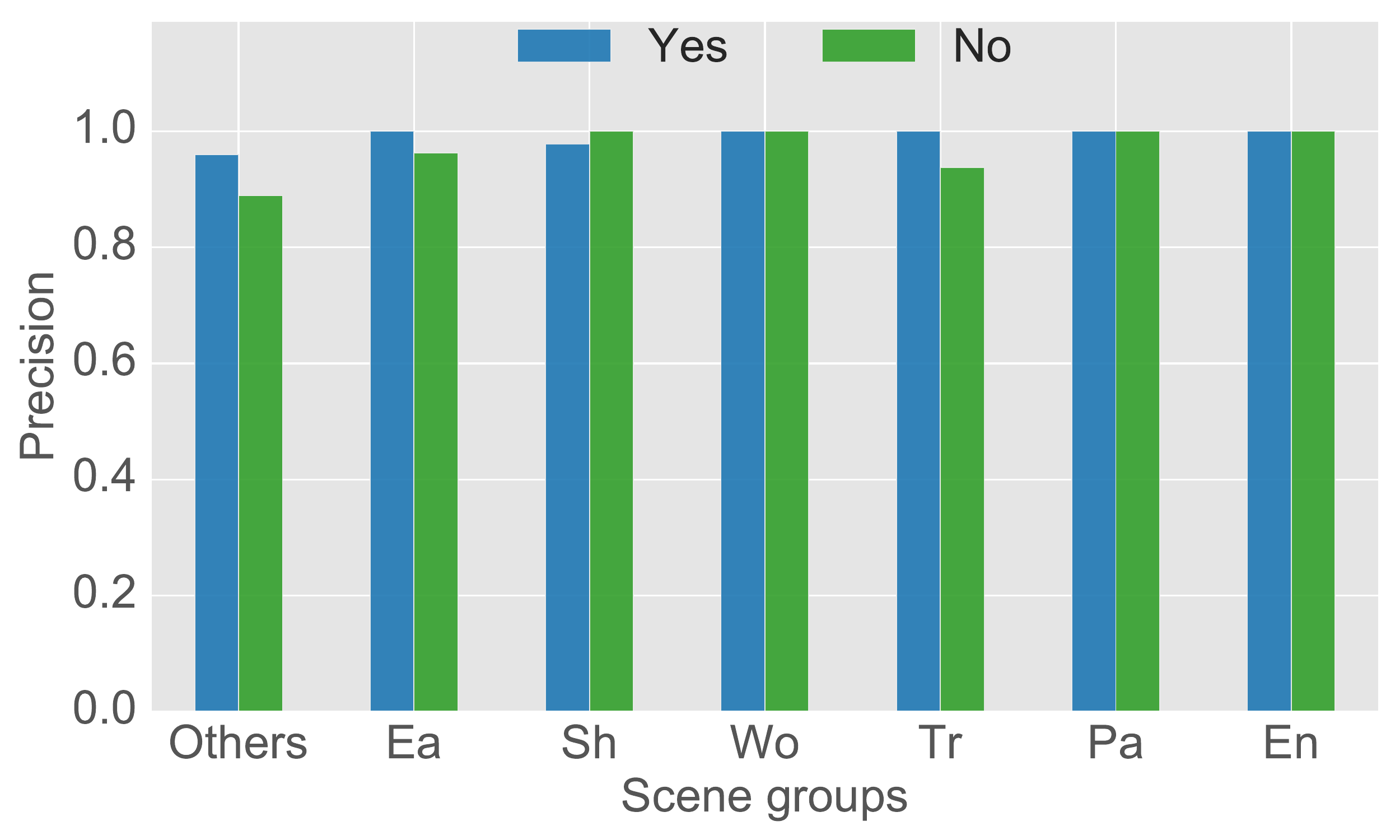}
  \label{fig:handprecision}
  }
\subfigure[Recall for different hand region sizes]{
  \includegraphics[width=0.31\textwidth]{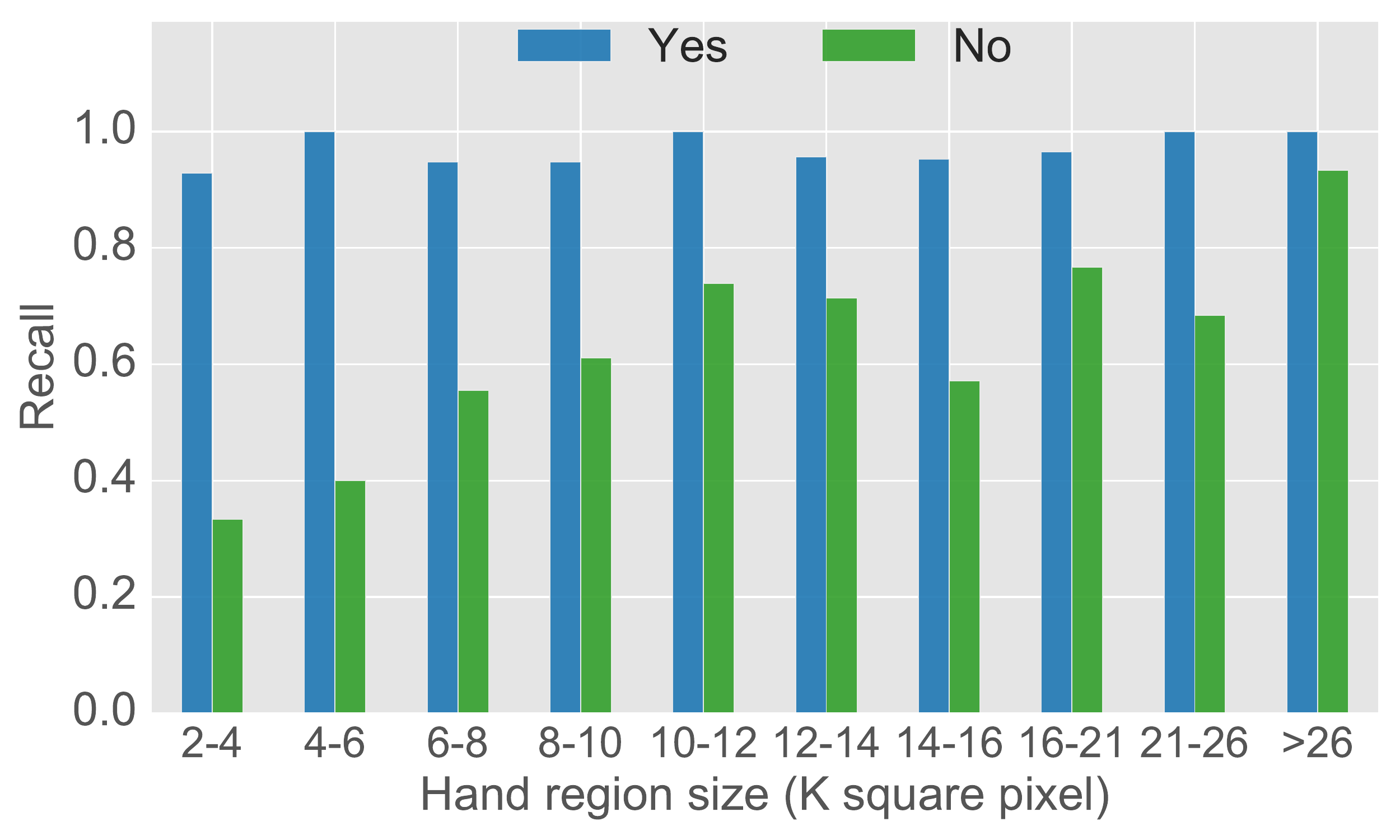}
  \label{fig:handrecallsize}
  }
\caption{Hand gesture recognition results.}
\label{fig:hand-recognition}
\end{figure*}

We asked our volunteers to take pictures for each other with different distances, angles, lighting conditions, and backgrounds.
We manually annotated all images with hand regions and the scene group it belongs to.
In total, we got $338$ hand gesture images with $208$ ``Yes'' gestures, $211$ ``No'' gestures, and $363$ natural hands.
The images covers $6$ out of $9$ scene groups.
For images do not belong to our summarized scene groups, we categorize them into group \textit{Others}.

Figure~\ref{fig:handcall} and Figure~\ref{fig:handprecision} show the recall and precision for ``Yes'' and ``No'' gestures under different scene groups.
The recall and precision of ``Yes'' gesture, the precision of ``No'' gesture reach $1.0$ for most of the scene groups, while the recall of ``No'' gesture achieves $0.70$ on average.
For \textit{Entertainment (En)}, the recall is low, resulting from dim illumination in most of the images we tested.
In general, the performance of gesture recognition does not show any marked correlation with different scene groups.
Therefore, we further investigated the recall in terms of gesture size, as low recall will greatly threatens user's privacy compared with precision.

Figure~\ref{fig:handrecallsize} plots the recall of gestures with varying hand region sizes.
Each image will be resized to about $800 \times 600$ square pixels, while keeping its aspect ratio.
We classify them into $10$ size intervals as plotted along the \textit{x}--axis.
The result shows ``Yes'' gestures can achieve more than $0.9$ recall for all sizes of hand region.
On the other hand, recall of ``No'' gestures tends to rise with increasing hand region size in general.
It indicates that the performance of ``No'' gesture recognition can be improved with more training data with smaller sizes.

In summary, the performance of gesture recognition demonstrates the feasibility of integrating gesture interaction in Cardea for flexible privacy preference modification. It performs extremely well for ``Yes'' gesture recognition, and there is room for improvement of ``No'' gesture recognition with more training samples.

\subsection{Overall Performance of Cardea Privacy Protection}

\begin{table}[tb]
\centering
\caption{Cardea's overall privacy protection performance.}
\label{tab:energy}
\begin{tabularx}{.485\textwidth}{lr|lr}
\toprule
\textbf{Overall accuracy}	& $\mathbf{86.4\%}$	& Protection accuracy	& $80.4\%$	\\
							& 					&  No protection accuracy& $91.0\%$	\\ \midrule
Face recognition accuracy	& $98.5\%$	& ``Yes'' gesture recall		& $97.9\%$	\\
scene classification recall	& $77.7\%$	& ``No'' gesture recall		& $77.3\%$	\\ \bottomrule

\end{tabularx}
\end{table}

After evaluating each vision task separately, we now present Cardea's overall privacy protection performance.
Faces in the image taken using Cardea end up being protected (e.g., blurred) or remain unchanged, correctly or incorrectly, depending on protection decisions made based on both user's privacy profile and results from vision tasks.
Therefore, we asked $5$ volunteers to register as Cardea users and set their privacy profiles.
Now the face recognition model is trained using $1100$ face feature vectors from $55$ people, including $50$ people from LFW dataset.
We take about $300$ images and get processed images.
As we focus more on face recognition rather than face detection, we only keep images that faces have been successfully detected.
In total we got $224$ images for evaluation.

Table~\ref{tab:overall} shows the final privacy protection accuracy, as well as performance of each vision task.
The protection accuracy shows $80.4\%$ faces that require protection are actually protected, and $91.0\%$ faces that do not ask for protection remain unchanged.
Overall, $86.4\%$ faces are processed correctly, though the scene classification recall and ``No'' gesture recall do not reach $80\%$.
The reason is that protection decision making process of Cardea sometimes can make up for mistakes happening in the early step.
For example, if user's ``No'' gesture is not detected, his face can still be protected when the predict scene is selected in user's profile.

In summary, Cardea achieves over $85\%$ accuracy for users in the real world.
Improvements of each vision part will directly benefit Cardea's overall performance in the future.

\subsection{Runtime and Energy Consumption}
We validate the client side implementations on Samsung Galaxy Note 4\footnote{\url{http://www.gsmarena.com/samsung_galaxy_note_4-6434.php}}, with 4$\times$2.7 GHz Krait 450 CPU, Qualcomm Snapdragon 805 Chipset, 3GB RAM, and 16 MP, f/2.2 Camera.
The server side is configured with Intel i7--5820K CPU, 16GB RAM, GeForce 980Ti Graphic Card (6GB RAM).
The client and server communicate via a TCP over Wi--Fi connection.

\begin{figure}[tb]
\centering
\includegraphics[width=0.9\columnwidth]{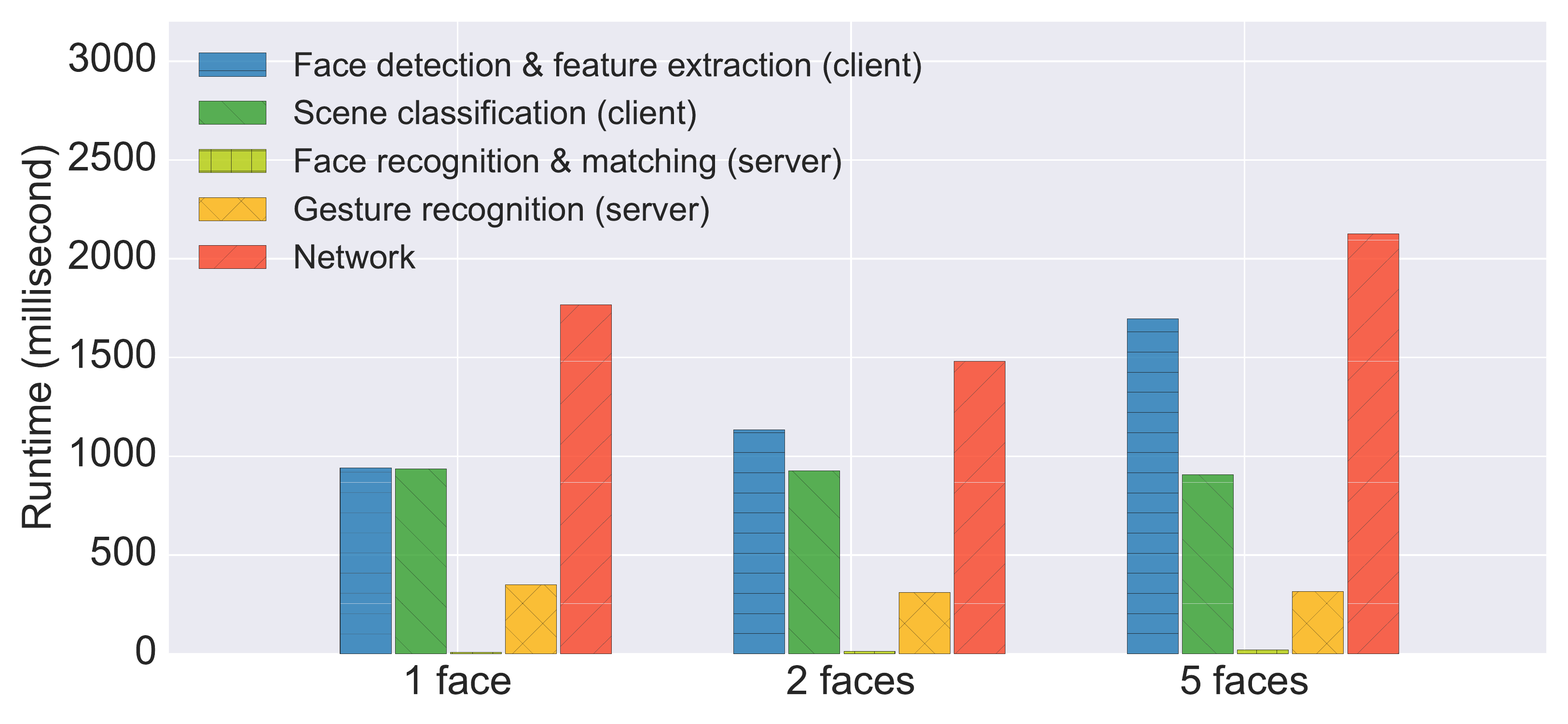}
\caption{Task level runtime of Cardea.}
\label{fig:runtime}
\end{figure}

Figure~\ref{fig:runtime} plots the time taken for Cardea to complete different vision tasks.
We take images with $1$ face, $2$ faces, and $5$ faces, in the size of $1920 \times 1080$.
The images will be compressed in JPEG format.
On average, the data sent is about $950$ KB.
Note, some vision tasks will not be triggered in some situations according to the decision workflow as explained in Section~\ref{sec:implementation}.
For example, if no user is recognized in the image, all other tasks will not start.
For the purpose of measurement, we still activate all tasks to illustrate the runtime difference of images captured with varying number of faces.

Among all vision tasks, face processing (i.e., face detection and feature extraction) and scene classification are performed on the smartphone.
The face processing takes about $900$ milliseconds for 1 face, and increases about $200$ milliseconds per additional face.
Therefore, it reaches about $1100$ and $1700$ milliseconds for $2$ faces and $5$ faces respectively.
This is the most fundamental step that runs on the smartphone locally due to privacy considerations.
Owing to the real--time OpenCV face detection implementation and lightened CNN feature extraction model, it takes less then $1/3$ overall runtime.
The scene classification takes around $900$ milliseconds per image, as it only performs once, independent of number of people in the image.
The face recognition and matching tasks on the server side take less than $30$ milliseconds for five people.
Though the time grows with increasing number of people, compared with other tasks they barely affect the overall runtime.
The gesture recognition also runs on the server, and it takes about $330$ milliseconds.
Similar to scene recognition, it performs on the whole image once regardless the number of faces.
According to the measurement, we find the network transmission accounts for a majority of the overall runtime due to the unstable network environment.

In general, photographers using Cardea to take pictures can get one processed image within $5$ seconds in the most heavy case (i.e., there is registered user who enables gestures, and scene classification will be triggered on the smartphone).
Compared with existing image capture platforms that also provide privacy protection such as I-pic \cite{aditya2016pic}, Cardea offers more efficient image processing functionality.

\begin{table}[tb]
\centering
\caption{Energy consumption of Cardea with different number of faces.}
\label{tab:overall}
\begin{tabularx}{.45\textwidth}{lrrr}
\toprule
			& Face recognition		& Whole process (uAh)	& \# of images \\ \midrule
1 face      & $217.2$ (std $3.4$)	& $1134.5$ (std $45.9$)	& $\sim 2800$		\\	
2 faces     & $344.1$ (std $13.1$)	& $1276.7$ (std $113.8$)	& $\sim 2500$		\\
5 faces     & $692.6$ (std $36.5$)	& $1641.1$ (std $66.0$)	& $\sim 2000$		\\ \bottomrule
\end{tabularx}
\end{table}

Next, we measure the energy consumption of taking pictures with Cardea on Galaxy Note 4 phone using the Monsoon Power Monitor \cite{powermonitor}.
The images are also taken in size of $1920 \times 1080$ square pixels.
The first two columns of Table~\ref{tab:energy} show the energy consumption for the face processing part only, and for the whole process (i.e., from taking a picture to getting the processed image) with 1 face, 2 faces, and 5 faces respectively.
The screen stays on during the whole process, therefore a large portion of the energy consumption is due to the always--on screen.
Moreover, we can observe that face processing energy is linear to face numbers.
All the other parts including scene classification, sending and receiving data are independent of the number of faces in the image, which is consistent with runtime measurements.

Using energy measurements, we also show Cardea's capacity on Galaxy Note 4 in the last column of Table~\ref{tab:energy}.
This device has a $3220$ mAh battery, therefore can capture about $2000$ high quality images with $5$ faces using Cardea.

\section{Related Work}
\label{sec:related-work}

Visual privacy protection has attracted extensive attentions these years due to increasing popularity of mobile and wearable devices with built--in cameras.
As a result, some research works have proposed solutions to address these privacy issues.

Halderman et al. \cite{halderman2004privacy} propose an approach in which closed devices can encrypt data together during recording utilizing short range wireless communication to exchange public keys and negotiate encryption key.
Only by obtaining all of the permissions from people who encrypt the recording can one decrypts it.
Juna et al. \cite{jana2013scanner, jana2013enabling} present methods that third--party applications such as perceptual and augmented reality applications have access to only higher-level objects such as a skeleton or a face instead of raw sensor feeds.
Raval et al. \cite{raval2016you} propose a system that gives users control to mark secure regions that the camera have access to, therefore cameras are prevented from capturing sensitive information.
Unlike above solutions, our work focuses on protecting bystanders' privacy by respecting their privacy preferences when they are captured in photos.

To identify individuals who request privacy protection, advanced techniques have been applied.
PriSurv \cite{chinomi2008prisurv} is a video surveillance system that identifies objects using RFID--tags to protect the privacy of objects in the video surveillance system.
Courteous Glass \cite{jung2014courteous} is a wearable camera integrated with a far-infrared imager that turns off recording when new persons or specific gestures are detected.
However, these approaches require extra facilities or sensors that are not currently equipped on devices.
Our approach takes advantage of state--of--the--art computer vision techniques that are reliable and effective.

Some other efforts rely on visual markers.
Respectful Cameras \cite{schiff2009respectful} is a video surveillance system that requires people who wish to remain anonymous wear colored markers such as hats or vests, and their faces will be blurred.
Bo et al. \cite{bo2014privacy} use QR code as privacy tags to link an individual with his photo sharing preferences, which empowers photo service providers to exert privacy protection following users' policy expressions, to mitigate the public's privacy concerns.
Roesner et al. \cite{roesner2014world} propose a general framework for controlling access to sensor data on continuous sensing platforms, allowing objects to explicitly specify access policies.
For example, a person can wear a QR code to communicate her privacy policy, and a depth camera is used to find the person.
Unfortunately, using existing markers like QR code is technically feasibly, but lacks usability, as few would wear QR code or other conspicuous markers in public places.
Moreover, static policies cannot satisfy people's privacy preferences all the time, which is context--dependent, varying from each other and changing from time to time. Therefore, we allow individuals to update their privacy profiles at anytime, providing other cues to inform cameras of necessary protection.

A recently interesting work I--pic \cite{aditya2016pic} allows people to broadcast their privacy preferences and appearance information to nearby devices using BLE.
This work can be incorporated into our framework.
Furthermore, we specify context elements that have not been considered before, such as scene and presence of others.
Besides, we provide a convenient mechanism for people to temporarily change their privacy preferences using hand gestures when facing the camera, while broadcasted data in I-pic may not be received by people who take images, or the received data is outdated.

\section{Conclusion and Future Work}
\label{sec:conclusion}

In this work, we designed, implemented, and evaluated Cardea, a visual privacy protection system that aims to address individuals' visual privacy issues caused by pervasive cameras.
Cardea leverages the state--of--the--art CNN models for feature extraction, visual classification and recognition.
With Cardea, people can express their context--dependent privacy preferences in terms of location, scenes, and presence of other people in the image.
Besides, people can show ``Yes'' and ``No'' gestures to dynamically modify their privacy preferences.
We demonstrated performances of different vision tasks with pictures ``in the wild'', and overall it can achieve about $86\%$ accuracy on users' requests.
We also evaluated runtime and energy consumed with Cardea prototype, which proves the feasibility of running Cardea for taking pictures while respecting bystanders' privacy preferences.

Cardea can be enhanced in different aspects.
The future work includes improving scene classification and hand gesture recognition performances, compressing CNN models to reduce overall runtime, and integrating Cardea with camera subsystem to enforce privacy protection measure.
Moreover, protecting people's visual privacy in the video is another challenging and significant task, which will make Cardea a complete visual privacy protection framework.


%
%



\bibliographystyle{IEEEtran}
\bibliography{references}

\end{document}